# Accelerating Plane Symmetric Cosmological Model with Bulk Viscous and Cosmic Strings in Lyra's Geometry

M. Krishna[1], SobhanBabu Kappala[2], R. Santhikumar[3]
[1]Raghu Institute of Technology, Visakhapatnam, Andhrapradesh-India,
Email: mandala.krishna.phd@gmail.com
[2]University Collage of Engineering, Narasaraopeta, AndhraPradesh-India,
Email: ksobhanjntu@gmail.com
[3]Aditya Institute of Technology and Management, Tekkali, Srikakulam Dist- AndhraPradesh-India
Email: skrmahanthi@gmail.com

**Abstract:** The present study deals with Lyra's geometry in plane symmetric metric discussed in the presence of Bulk viscous fluid and one-dimensional strings are assumed to be loaded with particles and the particle energy density. The variation of Hubble's parameter gives a constant value of decelerating parameter. The exact solution has been found for the plane-symmetric model in Lyra's geometry in the framework of bulk viscosity and string cosmology. Also, the bulk viscous pressure is assumed to be proportional to the energy density. The physical and geometrical properties of the model are also discussed.

**Key Wards** : Accelerated expansion, Bulk Viscosity, Cosmic string, Lyra geometry.

## 1.Introduction

At present there has been lost of interest in the study of cosmological models of the universe it creates more interest to understand the mystery of early stage structure formation of the universe. In the universe the role of inflationary phase is very important to solve the number of outstanding problems in cosmology like uniformity and static model of the universe.

To understand the uniformity and stat model of the universe, Einstein's field equations allow only non-static cosmological models for nonzero energy density. Hence, Einstein's general theory of relativity can be described in the form of geometry. Many modifications have been proposed in Riemannian geometry. In 1981, Weyl proposed a generalized Riemannian geometry that shows uniform gravitation and electromagnetism of the universe. So, it supports Einstein's proposed static model of the universe. Later Folland proposed a global formation of Weyl's geometry. In 1951, Lyra introduced a gauge function which is the modification of Riemannian geometry into a structureless manifold, it causes cosmological constants contained naturally from the geometry of the universe, Later Sen(1957) and Sen&Dumn(1971) formulated a new scalar tensor theorm in the form of Lyra's geometry, Hallford(1972) studied scalar-tensor theory in Lyra's geometry.

In the early stage of the universe there may be phase transition and spontaneous symmetric breaking giving us to rise a random network of stable line-like topological defects known as cosmic strings. It is well known that massive strings serve as seeds for large structures like galaxies and the cluster of galaxies in the universe. And also, many authors Letelier (1983), Krori et. al(1990), Reddy el.al(2003), Tripathy et.al(2008,2009) investigated different types of string cosmological theories in the frame work of general relativity. String cosmological models have played a vital role in the attention of the research work because of more importance of string theory in the early stage of evaluation of the universe.

Now a day there are unpredictable problems with negative pressure causing repulsive gravity popularly called an inflationary phase getting the accelerated expansion of the universe. Bulk viscosity contributes a negative pressure causing the accelerating expansion of the universe. Hence there is a lot of importance to study the bulk viscosity of the universe. Bulk viscous cosmological models in general relativity have been studied by several authors Bali et.al.(2007) and Rao et. al(2011,2012). Reddy et.al (2013) studies Kaluza-Klein universe with cosmic strings and bulk viscosity in a scalar-tensor theory of gravitation. Naidu et. al(2012) discussed the LRS-Bianchi type-II model with cosmic strings and bulk viscosity in a scalar-tensor theory of gravitation. Vijayasanthi et.al (2017) studied bulk viscous string cosmological models in f(R) gravity. Katore et.al (2018) has been investigated bulk viscous string cosmological models in f(G) gravity.

Recently several authors studied Lyra's geometry in different cosmological models. Reddy (2005) developed plane symmetric cosmic string theory in Lara manifold.Adhav et.al (2009) proposed a zero-mass scalar field with bulk viscous cosmological solutions in Lyra's geometry. Shri Ram et. al(2010) studied the Bianchi type-V cosmological model in Lyra's geometry in the presence of perfect fluid. Asgar and Ansari(2014) studied the Bianchi $VI_0$ bulk viscous cosmological model in Lyra's geometry. Sing and Sarma(2014) studied the anisotropic Dark energy Bianchi type-II cosmological model in Lyra's geometry. Kalyani and Sanjit(2017) studied FRW cosmological metric in Lyra's geometry in the presence of Particle creation. Dinesh and Rashid(2019) studied the Brans-Dicke scalar filed cosmological model in Lyra's geometry. Bakry et.al (2022) studied particle creation and big rip cosmology in Lyra's geometry. From the motivation of all the above studied in this paper we studied Lara's geometry in plane symmetric metric and the energy-momentum tensor consisting of Bulk viscous fluid is one-dimensional strings are assumed to be loaded with particle and the particle energy density.

## 2. Metric and field Equations :

Field equation of plane symmetric metric is

$$ds^2 = dt^2 - A^2(dx^2 + dy^2) - B^2 dz^2 \qquad (1)$$

Where A,B,C are functions

of cosmic time t.

The Einstein modified filed equation in normal gauge for Lyra's modified obtained by Sen(1957) given by (where $8\pi G = 1, C = 1$)

$$R_i^j - \frac{1}{2}g_i^j R + \left[\frac{3}{2}\phi_i\phi^j - \frac{3}{4}\phi_k\phi^k g_i^j\right] = -T_i^j \tag{2}$$

Where $\phi_i$ is the displacement vector defined as $\phi_i = (\beta(t), 0,0,0)$ \hfill (3)

Let $T_{ij}$ is the energy- momentum tensor of the matter and comma and semicolon denoted partial and covariant differentiation .

Also, $T^{ij}_{;j} \equiv 0$ \hfill (4)

We Assume the cosmic matter consisting Bulk viscous fluid the energy- momentum tensor is

$$T_{ij} = (\rho + \bar{p})u_i u_j - \bar{p}g_{ij} - \lambda x_i x_j \tag{5}$$

Where $\rho$ is rest energy density of the system

$\lambda$ is tension in the string and

The total pressure contains a bulk viscous pressure $\xi$ and hubbles parameter H is

$$\bar{p} = (p - 3\xi H) \tag{6}$$

Here $u^i = \delta_0^i$ is the four velocity vector and $x^i$ is spacelike vector represented the anisotropic direction of the string,

Hence $u^i$ and $x^i$ satisfies the equation $g_{ij}u^i u_j = 1, g_{ij}x^i x_j = 1, u^i x_i = 0$ \hfill (7)

The one dimensional strings are assumed to be loaded with particle and the particle energy density is

$$\rho_p = (\rho - \lambda) \tag{8}$$

where $\rho, \lambda, \bar{p}, \phi$ are functions of cosmic time t.

We have $T_{11} = T_{22} = -\bar{p}$ \quad $T_{33} = -(\bar{p} + \lambda)$, and $T_{00} = \rho$ \hfill (9)

For the equations of the metric. (1) , the field equation eq.(2) together with eq.(3) yields that

$$\frac{\ddot{A}}{A} + \frac{\ddot{B}}{B} + \frac{\dot{A}\dot{B}}{AB} + \frac{3}{4}\beta^2 = -\bar{p} \tag{10}$$

$$2\frac{\ddot{A}}{A} + \left(\frac{\dot{A}}{A}\right)^2 + \frac{3}{4}\beta^2 = -(\bar{p} + \lambda) \tag{11}$$

$$\left(\frac{\dot{A}}{A}\right)^2 + \frac{\dot{A}\dot{B}}{AB} - \frac{3}{4}\beta^2 = \rho \tag{12}$$

The overhead dot denotes ordinary differentiation with respect to t.

Eq.(4) leads that

$$\dot{\rho} + \left(2\frac{\dot{A}}{A} + \frac{\dot{B}}{B}\right)(\rho + \bar{p}) + \lambda\left(\frac{\dot{B}}{B}\right) = 0 \tag{13}$$

And conservation of L.H.S of eq.(2) leads that

$$R_i^j - \frac{1}{2}g_i^j R + \left[\frac{3}{2}\phi_i\phi^j - \frac{3}{4}\phi_k\phi^k g_i^j\right] = 0 \tag{14}$$

$$\frac{3}{2}\phi_i\left[\frac{\partial\phi^j}{\partial x^j} + \phi^l\Gamma_{lj}^j\right] + \frac{3}{2}\phi^j\left[\frac{\partial\phi_i}{\partial x^j} - \phi_l\Gamma_{ij}^l\right] - \frac{3}{4}g_i^j\phi_k\left[\frac{\partial\phi^k}{\partial x^j} + \phi^l\Gamma_{lj}^k\right] - \frac{3}{4}g_i^j\phi^k\left[\frac{\partial\phi_k}{\partial x^j} - \phi_l\Gamma_{kj}^l\right] = 0 \tag{15}$$

Eq.(15) leads that

$$\frac{3}{2}\beta\dot{\beta} + \frac{3}{2}\beta^2\left(2\frac{\dot{A}}{A} + \frac{\dot{B}}{B}\right) = 0 \tag{16}$$

The spatial volume and average scale factor for the metric (1) is defined as

$$V = R^3 = A^2 B \tag{17}$$

The generalized mean Hubbles parameter H for the metric (1) id defined as

$$H = \frac{\dot{R}}{R} = \frac{1}{3}\left(2\frac{\dot{A}}{A} + \frac{\dot{B}}{B}\right) = \frac{1}{3}\left(\frac{\dot{V}}{V}\right) \tag{18}$$

Where $H_x = H_y = \frac{\dot{A}}{A}$ and $H_z = \frac{\dot{B}}{B}$.

The scalar expansion $\theta$ for the metric (1) is defined as

$$\theta = \left(2\frac{\dot{A}}{A} + \frac{\dot{B}}{B}\right) \tag{19}$$

The sear scalar $\sigma$ for the metric (1) is defined as

$$\sigma^2 = \frac{\theta^2}{6} = \frac{3H^2}{2} = \frac{\left(2\frac{\dot{A}}{A} + \frac{\dot{B}}{B}\right)^2}{6} \tag{20}$$

The mean anisotropy parameter $A_m$ is given by

$$A_m = \frac{1}{3}\sum_{i=1}^{3}\left[\frac{\Delta H_i}{H}\right]^2 \tag{21}$$

Where $\Delta H_i = H_i - H (i = 1,2,3)$

An important observational quantity in cosmology is the decelerating parameter q which is defined as

$$q = -\frac{R\ddot{R}}{\dot{R}^2} = -\left(\frac{\dot{H}+H^2}{H^2}\right) \tag{22}$$

## 3. Solutions and Models

The field equations (10)-(12) are a system of three independent equations in six unknowns $A, B, \bar{p}, \rho, \beta,$ and $\lambda$. Also the field equations are highly non-linear in nature and therefore we require the following plausible physical conditions:

(i) A more general relation between the proper rest energy density $\rho$ and string tension density $\lambda$ is taken to be

$$\rho = r\lambda \qquad (23)$$

Where $r$ is an arbitrary constant that can take both negative and positive values. The negative value of r leads to the absence of string in the universe and positive value shows the presence one dimensional strings in the cosmic field. The energy density of the particles attached to the string is

$$\rho_p = \rho - \lambda = (r-1)\lambda \qquad (24)$$

(ii) For a barotropic fluid, the combined effect of the proper pressure and the barotropic bulk viscous pressure can be expressed as

$$\bar{p} = p - 3\xi H = \varepsilon\rho \qquad (25)$$

Where $\varepsilon = \varepsilon_0 - \zeta$ and $p = \varepsilon_0\rho \ (0 \leq \varepsilon_0 \leq 1)$

(iii) Since the shear scalar $\sigma$ is proportional to scalar expansion $\theta$

Clearly $\sigma \propto \theta$ means that $\lim_{n \to \infty} \neq 0$

The model is anisotropic in nature, by Collin et.al(1980)

We can consider $A = B^n$ for $n \neq 1$ , (26)

the model represents anisotropic model.

Using the above conditions the field equations (10)-(11) can be reduces to the equation

$$a\frac{\ddot{B}}{B} + b\left(\frac{\dot{B}}{B}\right)^2 = n\frac{\dot{B}}{B} \qquad (27)$$

Where $a = [r(1-\varepsilon)(1-n) - (n+1)], b = [2r(1-\varepsilon)(n-n^2) - (n^2+n)]$

By solving eq.(22) we have

$$A = k_2{}^n \exp\left(\frac{n^2 t}{a+b}\right), \quad B = k_2 \exp\left(\frac{nt}{a+b}\right) \qquad (28)$$

Where $k_2 = \left[\left[\frac{a+b}{n}\right]k_1\right]^{\left(\frac{a}{a+b}\right)}$ and $k_1$ is constants of integration.

The filed equation (1) reduces to

$$ds^2 = dt^2 - k_2{}^n \exp\left(\frac{n^2 t}{a+b}\right)(dx^2 + dy^2) - k_2 \exp\left(\frac{nt}{a+b}\right) dz^2 \qquad (29)$$

By using a suitable transformation of coordinate t=T, the above equation is transform to

$$ds^2 = dT^2 - k_2{}^n \exp\left(\frac{n^2 t}{a+b}\right)(dx^2 + dy^2) - k_2 \exp\left(\frac{nt}{a+b}\right) dz^2 \tag{30}$$

## 4. The Physical and Geometrical Features

Equation .30 is the plane symmetric model with cosmic strings and bulk viscosity in Lyra's geometry with the following physical and kinematical parameter which are very important for a physical insight of the model.

The displacement vector $(\beta)$,

$$\beta = k_3 \exp\left(\frac{-2n^2 - n}{a+b}\right) t \tag{31}$$

Where $k_3 = c_1 k_2^{-(2n+1)}$, here $c_1$ is integration constant

The energy density $(\rho)$,

$$\rho = k_4 - \frac{3k_3^2}{4} \exp\left(\frac{-4n^2 - 2n}{a+b}\right) t \tag{32}$$

Where $k_4 = (a+b)^2 \left(\frac{n k_2^{(n+1)} + k_2^{2n}}{n^4}\right)$

Proper pressure and the barotropic bulk viscous pressure

$$\bar{p} = (\varepsilon_0 - \zeta)\rho = (\varepsilon_0 - \zeta)\left[k_4 - \frac{3k_3^2}{4}\exp\left(\frac{-4n^2 - 2n}{a+b}\right) t\right] \tag{33}$$

$$p = \varepsilon_0 \rho = \varepsilon_0 \left[k_4 - \frac{3k_3^2}{4}\exp\left(\frac{-4n^2 - 2n}{a+b}\right) t\right] \quad , for\ (0 \leq \varepsilon_0 \leq 1) \tag{34}$$

Tension of the string $(\lambda)$

$$\lambda = \frac{1}{r}\rho = \frac{1}{r}\left[k_4 - \frac{3k_3^2}{4}\exp\left(\frac{-4n^2 - 2n}{a+b}\right) t\right] \tag{35}$$

The energy density of the particles attached to the string is

$$\rho_p = \rho - \lambda = \left(r - \frac{1}{r}\right)\left[k_4 - \frac{3k_3^2}{4}\exp\left(\frac{-4n^2 - 2n}{a+b}\right) t\right] \tag{36}$$

It is observed that the displacement vector $\beta$ is very large at the time of evaluation of the universe and decreases rapidly with the evaluation of the model analogous to cosmological constant $(\Lambda)$. The energy density $(\rho)$ and Pressure $(p)$, the tension of the string $(\lambda)$, particle energy density $\rho_p$ all are decreasing with respect to cosmic time. Hence, here we observed that the decreasing pressure undergoes negative pressure given that the universe is expanding called as inflation phase. Because of the decrease in the nature of cosmic string with respect to cosmic time a topological phase transition may be observed during the evaluation of the universe.

Spatial volume is

$$V = k_3 \exp\left(\frac{2n^2+n}{a+b}\right) t \tag{37}$$

The average scale factor

$$R = \left(k_3 \exp\left(\frac{2n^2+n}{a+b}\right) t\right)^{\frac{1}{3}} \tag{38}$$

Also,

$$\lim_{t\to\infty} \frac{\sigma^2}{\theta^2} = \frac{1}{6} \neq 0 \tag{39}$$

Thus the model does not represent Isotropic for large values of t.

The Coefficient of Bulk Viscoussity ($\xi$) is

$$\xi = \left(\frac{(a+b)\zeta}{2n^3+n^2}\right)\left[k_4 - \frac{3k_3^2}{4}\exp\left(\frac{-4n^2-2n}{a+b}\right) t\right] \tag{40}$$

The decelerating parameter for this model is obtained by

$$q = -1 \tag{41}$$

By recent study (Permutter et.al.(1997,1998); Tonry(2003);John(2004); Knop et .al.(2003) informed that if the range of $-1 \leq q < 0$ and the present-day universe is undergoing an accelerated expansion. Here, the model obtained q= -1 indicates that the universe is accelerating at present. Since decelerating parameter becomes -1 indicates that the model is expanding with constant velocity.

## 5.Conclusion :

In this paper, we obtain a plane-symmetric type in the framework of Lyra's geometry for a time-dependent displacement field when the energy-momentum tensor is a viscous fluid containing one-dimensional string. Clearly it is observed that the displacement vector ($\beta$) is proportional to energy density ($\rho$), hence Proper pressure(p), the energy density ($\rho$), one dimensional tension of the string ($\lambda$), the Coefficient of Bulk Viscosity ($\xi$) and the energy density of the particles attached to the string $\rho_p$ are very high at the during late cosmic time of evaluation and expansion of universe . And also all the above components are very low during at early stage universe. It is also obsevered that, Since $\lim_{t\to\infty} \frac{\sigma^2}{\theta^2} = \frac{1}{6} \neq 0$ , the model does not approach isotropic for large values of t. Also , Since $q = -1$ indicates that the universe is accelerating at present. It is observed that the decreasing pressure undergoes to negative pressure gives that the universe is expanding rapidly called as inflation phase. Because of the decrease in the nature of cosmic string with respect to cosmic time a topological phase transition may be observed during the evaluation of the universe for this model. It is observed that the string model in the static case gives singularity of the universe. Finally, here we presented a plane symmetric bulk viscous cosmic string model in Lyra's geometry.